\documentclass[]{research}
\usepackage{amsmath}
\usepackage{amsfonts}
\usepackage{hyperref}

\usepackage{url}
\usepackage{booktabs}
\usepackage{multirow}
\usepackage{graphicx}
\usepackage{array}
\usepackage{xspace}
\usepackage{algorithm}
\usepackage{algpseudocode}
\usepackage{siunitx}
\usepackage{tikz}
\usetikzlibrary{arrows.meta,positioning,calc}

\crefname{figure}{Figure}{Figures}
\crefname{table}{Table}{Tables}
\crefname{section}{Section}{Sections}
\crefname{algorithm}{Algorithm}{Algorithms}
\crefname{lstlisting}{Listing}{Listings}

\usepackage{listings}
\usepackage{xcolor}
\lstdefinestyle{mkcode}{
  basicstyle=\ttfamily\footnotesize,
  columns=fullflexible,
  breaklines=true,
  frame=single,
  rulecolor=\color{black!20},
  frameround=tttt,
  showstringspaces=false,
  tabsize=2,
  keywordstyle=\color{blue!60!black},
  commentstyle=\color{green!40!black},
  stringstyle=\color{orange!60!black},
  xleftmargin=2pt,
  xrightmargin=2pt,
}

\lstdefinelanguage{CUDA}{
  language=C++,
  morekeywords={__global__,__host__,__device__,__shared__,__align__,__launch_bounds__,
                __syncthreads,__threadfence_system,__nanosleep,uint32_t,uint64_t,uint16_t,uint8_t,
                constexpr,auto},
}

\definecolor{mkorange}{HTML}{EB6834}
\definecolor{mkblue}{HTML}{2A78D6}
\definecolor{mkaqua}{HTML}{1BAF7A}
\definecolor{mkviolet}{HTML}{4A3AA7}
\definecolor{mkgray}{HTML}{52514E}

\newtcolorbox{takeaway}{enhanced, colback=mkorange!6, colframe=mkorange!80!black,
  boxrule=0pt, leftrule=2.5pt, arc=0pt, outer arc=0pt, left=7pt, right=7pt, top=4pt, bottom=4pt,
  fontupper=\small, before skip=8pt, after skip=10pt}

\newcommand{\system}{\textsc{mKernel}\xspace}

\title{\fontsize{20}{22}\selectfont \system: Fast Multi-GPU, Multi-Node Fused Kernels}

\author[1]{Ziming Mao}
\author[2]{Yihan Zhang}
\author[3]{Shawn Wei Chew}
\author[2]{Shuang Ma}
\author[4]{Costin Raiciu}
\author[2]{Yang Zhou}
\author[1]{Scott Shenker}
\author[1]{Ion Stoica}

\affiliation[1]{UC Berkeley}
\affiliation[2]{UC Davis}
\affiliation[3]{UCLA}
\affiliation[4]{University Politehnica of Bucharest}

\abstract{Communication has become a bottleneck in distributed training and inference of large models.
Overlapping communication with computation at the granularity of kernels, on separate streams, reduces only part of this communication cost. Fused kernels often have better performance by transmitting each output tile as soon as it is produced, but existing fused kernels are largely confined to a single NVLink domain. We present \system, a library of \textbf{multi-GPU, multi-node fused kernels} that \textbf{overlap computation, intra-node NVLink communication, and inter-node RDMA at tile granularity}. \system partitions the streaming multiprocessors (SMs) of a persistent kernel into compute and communication roles, and an on-GPU controller \textbf{tunes the SM partition adaptively at run time}, since the best SM partition varies with the kernel and the input shape. It structures data movement hierarchically so that data traversing the inter-node network is minimized. Finally, it drives the network from the GPU through a lightweight command queue and host proxy implemented directly on RDMA verbs, which allows \textbf{the same kernels to run on any network backend} (e.g. InfiniBand and on AWS EFA); we observe, surprisingly, that GPUDirect Async (IBGDA) yields little additional benefit over host-assisted GPU-initiated communication. We implement five kernels spanning tensor, sequence, and expert parallelism. On two 16-GPU H200 clusters,
\system achieves speedups of up to $1.72\times$ on GEMM+AllReduce and $1.88\times$ on Ring Attention.}

\metadata[Code]{\href{https://github.com/uccl-project/mKernel}{https://github.com/uccl-project/mKernel}}
\metadata[Blog]{\href{https://uccl-project.github.io/posts/mkernel/}{https://uccl-project.github.io/posts/mkernel/}}

\begin{document}

\maketitle
\section{Introduction}
\label{sec:intro}

Training and serving large models requires partitioning either model layers or input sequence across many GPUs, using tensor, sequence, and expert parallelism~\cite{shoeybi2019megatron,liu2023ringattention,lepikhin2021gshard}. In production mixture-of-experts (MoE) training, communication accounts for 43.6\% of the forward pass and 32\% of end-to-end training time~\cite{jin2026megascalemoe}, and across popular MoE models and frameworks, inter-device communication accounts for up to 47\% of execution time~\cite{zhang2025comet}. The imbalance is widening because accelerator compute throughput is growing faster than network bandwidth: a GB300 NVL72 rack provides 720\,PFLOP/s of FP8 compute~\cite{nvidia_gb300}, yet each of its GPUs reaches devices outside the rack through a single NIC of 400--800\,Gb/s.

 Prior work overlaps computation and communication at different granularities. Intra-node fusion can
  overlap computation with local transfers while leaving inter-node communication to a separate collective
  (\cref{fig:timeline}(a)). Two-stream pipelines communicate completed chunks while computing subsequent
  chunks, but release each chunk only at its kernel boundary (\cref{fig:timeline}(b)). Tile-level
  fusion enables finer overlap by exposing tile readiness within the
  kernel~\cite{chang2024flux,zhang2025comet,sul2025parallelkittens}. Most existing works focus on \textit{single-node};  and many systems fuse only selected stages of computation, intra-node communication, and inter-node
  communication, leaving the remaining stages to execute separately. \system integrates computation, intra-node NVLink
  communication, and inter-node RDMA within a single fused kernel, coordinating tile-level progress across all
  three stages (\cref{fig:timeline}(c)).

\begin{figure}[t]
\centering
\resizebox{0.95\linewidth}{!}{%
\begin{tikzpicture}[x=1cm,y=0.8cm, font=\sffamily\small,
  blk/.style={rounded corners=1.5pt, text=white, font=\sffamily\scriptsize\bfseries, inner sep=0pt, minimum height=0.34cm},
  lab/.style={anchor=east, font=\sffamily\footnotesize, text=mkgray},
  ttl/.style={anchor=west, font=\sffamily\footnotesize\bfseries}]
\begin{scope}[shift={(0,5.8)}]
\node[ttl] at (0,1.95) {(a) Intra-node fusion, followed by an inter-node collective};
\node[lab] at (2.3,1.3) {Compute SMs};
\node[lab] at (2.3,0.65) {NVLink};
\node[lab] at (2.3,0) {Network};
\foreach \i in {0,...,5} {
  \node[blk, fill=mkorange, minimum width=1.0cm] at ({2.5+1.05*\i+0.5},1.3) {\i};
  \node[blk, fill=mkaqua, minimum width=0.25cm] at ({3.55+1.05*\i+0.125},0.65) {};
}
\node[blk, fill=mkblue, minimum width=3.0cm] at (10.6,0) {inter-node collective};
\draw[-{Stealth[length=2mm]}, black!50] (2.5,-0.4) -- (13.2,-0.4);
\draw[black!50] (12.1,-0.48) -- (12.1,-0.32);
\node[anchor=north, font=\sffamily\scriptsize, text=mkgray] at (12.1,-0.43) {end};
\end{scope}
\begin{scope}[shift={(0,2.9)}]
\node[ttl] at (0,1.95) {(b) Two-stream overlap at chunk boundaries};
\node[lab] at (2.3,1.3) {Stream 1: compute};
\node[lab] at (2.3,0.65) {Stream 2: NVLink};
\node[lab] at (2.3,0) {Stream 2: network};
\foreach \j/\tiles in {0/{0--1},1/{2--3},2/{4--5}} {
  \node[blk, fill=mkorange, minimum width=2.0cm] at ({2.5+2.2*\j+1.0},1.3) {\tiles};
  \node[blk, fill=mkaqua, minimum width=0.5cm] at ({4.6+2.2*\j+0.25},0.65) {};
  \node[blk, fill=mkblue, minimum width=1.0cm] at ({5.15+2.2*\j+0.5},0) {\tiles};
}
\draw[-{Stealth[length=2mm]}, black!50] (2.5,-0.4) -- (13.2,-0.4);
\draw[black!50] (10.55,-0.48) -- (10.55,-0.32);
\node[anchor=north, font=\sffamily\scriptsize, text=mkgray] at (10.55,-0.43) {end};
\end{scope}
\node[ttl] at (0,1.95) {(c) Intra- and inter-node fusion at tile granularity (\system)};
\node[lab] at (2.3,1.3) {Compute SMs};
\node[lab] at (2.3,0.65) {NVLink};
\node[lab] at (2.3,0) {Network};
\foreach \i in {0,...,5} {
  \node[blk, fill=mkorange, minimum width=1.0cm] at ({2.5+1.05*\i+0.5},1.3) {\i};
  \node[blk, fill=mkaqua, minimum width=0.25cm] at ({3.55+1.05*\i+0.125},0.65) {};
  \node[blk, fill=mkblue, minimum width=0.5cm] at ({3.85+1.05*\i+0.25},0) {\i};
}
\draw[-{Stealth[length=2mm]}, black!50] (2.5,-0.4) -- (13.2,-0.4);
\draw[black!50] (9.6,-0.48) -- (9.6,-0.32);
\node[anchor=north, font=\sffamily\scriptsize, text=mkgray] at (9.6,-0.43) {end};
\end{tikzpicture}}
\caption{\textbf{Three schedules for computation and communication.} (a) Intra-node fusion followed by a separate inter-node collective. (b) Two streams overlap communication of completed two-tile chunks with computation of subsequent chunks; each producer-kernel boundary releases a chunk. (c) \system schedules both communication tiers within the kernel, exposing tile-level readiness signal. We note that existing compute kernels already operate on tile granularity, so GEMM efficiency is not affected. Numbers identify output tiles; widths are schematic, not measured.}
\label{fig:timeline}
\end{figure}

Extending fusion across nodes is more difficult than fusing within a node. The inter-node network provides roughly one ninth of the per-GPU bandwidth of NVLink on our clusters, so the two tiers cannot be treated uniformly. The Network Interface Card (NIC) is a separate PCIe device with its own work queues, and GPUDirect Async requires NICs whose work queues the GPU can access directly. Transports also differ in their delivery guarantees: InfiniBand reliable connections deliver RDMA writes in order, whereas the Scalable Reliable Datagram (SRD) transport of AWS EFA~\cite{shalev2020srd} does not, so a completion flag written after the data may become visible before it. Both intra-node communication and inter-node communication require SMs, which compete with compute over shared resources.

This paper presents \system, a library of \textbf{inter-node fused kernels} in which computation, intra-node NVLink communication, and inter-node RDMA communication overlap \textbf{at tile granularity} (\cref{fig:timeline}(c)). Its design follows four principles. The first is \textbf{SM specialization}: each kernel assigns its thread blocks to compute, intra-node communication, inter-node send, and inter-node receive roles. The second is \textbf{hierarchical data movement}: data is locally reduced, or broadcasted over NVLink through NVSwitch, so that traffic traversing the inter-node network is minimized across pairs of GPUs across nodes. The third is \textbf{portable host-assisted GPU-initiated communication}: kernels enqueue compact transfer commands that a lightweight host proxy submits to the NIC as RDMA writes. This path is implemented directly on \texttt{libibverbs}, without NCCL or NVSHMEM, so the same kernels run on InfiniBand (ConnectX-7) and on AWS EFA. We also implemented GPUDirect Async (IBGDA) for ConnectX-7 and found little performance difference compared with host-assisted GPU-initiated communication. The fourth is \textbf{dynamic SM partitioning at runtime}: an on-GPU controller adjusts the compute--communication SM partition using measured progress and remaining work. 

We implement five kernels with \system. For tensor parallelism we provide AllGather+GEMM, GEMM+Reduce\-Scatter, and GEMM+All\-Reduce; for sequence parallelism, Ring Attention; and for expert parallelism, MoE Dispatch+GEMM. We evaluate them on two clusters of 2 nodes $\times$ 8 H200 GPUs, one cluster is interconnected with ConnectX-7 and the other cluster with EFA. Relative to cuBLAS or FlashAttention followed by NCCL, \system achieves speedups of up to $1.41\times$ on AllGather+GEMM, $1.72\times$ on GEMM+AllReduce, $1.88\times$ on Ring Attention. It also outperforms Triton-distributed, Flux, Mercury, MagiAttention, and ring-flash-attention in most configurations.

This paper makes three contributions. First, it analyzes the constraints that arise when fused kernels span multiple nodes (\cref{sec:background}). Second, it presents the design of \system, including a host-assisted GPU-initiated communication layer that supports both InfiniBand and AWS EFA (\cref{sec:design}), as well as one with IBGDA that runs on InfiniBand. Third, it describes five inter-node fused kernels (\cref{sec:kernels}) and evaluates them against seven baselines on two clusters (\cref{sec:eval}). \system is open source and available at \url{https://github.com/uccl-project/mKernel}.

\section{Background and Motivation}
\label{sec:background}

\subsection{Communication hierarchy of GPU clusters}

We evaluate two separate clusters: one uses InfiniBand between nodes, and the other uses AWS EFA (\cref{tab:hierarchy}). Both connect eight GPUs per node through NVLink and NVSwitch. The communication bandwidth across nodes is significantly lower within a node. ThunderKittens and ParallelKittens already provide the tile abstractions, asynchronous transfers, and intra-node communication primitives needed within a single node~\cite{spector2025thunderkittens,sul2025parallelkittens}. Across nodes, GPUDirect RDMA lets the NIC transfer payloads directly between GPU memories~\cite{gpudirectrdma}. The resulting kernel must allocate SMs to computation, local communication, and network coordination despite the different bandwidths of these paths.

\begin{table}[t]
\centering
\small
\caption{Communication hierarchy of two separate H200 clusters: one uses InfiniBand (IB), and the other uses AWS EFA. Both share the within-GPU and within-node hierarchy shown here. Crossing a node boundary changes which SMs coordinate data movement. NVLink and network bandwidths are per GPU, per direction.}
\label{tab:hierarchy}
\begin{tabular}{@{}l l r l l@{}}
\toprule
\textbf{Level} & \textbf{Link} & \textbf{Bandwidth} & \textbf{SM roles} & \textbf{Transfer mechanism} \\
\midrule
Within a GPU  & HBM3e              & 4.8\,TB/s & compute & loads/stores, TMA \\
Within a node & NVLink + NVSwitch  & 450\,GB/s & local communication & TMA, NVSwitch \\
Across nodes (IB)  & InfiniBand (CX7)   & 50\,GB/s & send and receive & NIC RDMA \\
Across nodes (EFA) & AWS EFA (SRD)      & 50\,GB/s & send and receive & NIC RDMA \\
\bottomrule
\end{tabular}
\end{table}

\subsection{Overlapping communication and computation in fused kernels}

Tile-level fusion achieves fine-grained compute-communication overlap inside the kernel: communication can consume each output tile when it is produced, and computation can consume each input tile when it arrives. Flux integrates communication into GEMM epilogues and readiness checks into GEMM prologues, including support for inter-node writes through NVSHMEM~\cite{chang2024flux}. Comet and MegaScale-MoE apply fine-grained overlap to MoE workloads~\cite{zhang2025comet,jin2026megascalemoe}. TileLink, Triton-distributed, and Mercury expose computation and communication through compiler primitives~\cite{zheng2025tilelink,zheng2025tritondistributed,guan2025mercury}.

\paragraph{SM allocation in a fused kernel.}

Consider a GPU with $S$ SMs executing a fused kernel that assigns $S_c$ of them to communication. If compute throughput scales approximately with the number of compute SMs and the three stages overlap in steady state, execution time can be approximated as
\begin{equation}
T_{\text{fused}} \;\approx\; \max\big(T_{\text{compute}}\cdot\tfrac{S}{S-S_c},\;\; T_{\text{NVLink}},\;\; T_{\text{network}}\big) \;+\; T_{\text{fill/drain}} \;+\; T_{\text{sync}}.
\label{eq:fused}
\end{equation}
Here, $T_{\text{compute}}$ is compute time with all $S$ SMs available; transfer times depend on the selected allocation. 
The final terms account for pipeline fill/drain and synchronization. \textbf{The central tradeoff is that additional communication SMs can reduce communication time while slowing computation.}
We use this observation to motivate adaptive allocation of SMs and hierarchical communication to minimize $T_{network}$.

\subsection{Challenges of designing efficient multi-GPU, multi-node fused kernels}
\label{sec:challenges}

Five challenges guide  design in \cref{sec:design}.

\paragraph{C1: Intra-node versus inter-node bandwidth asymmetry.}
The bandwidth gap between the NVLink domain and the inter-node network can make $T_{\text{network}}$ the dominant term in \cref{eq:fused}. This motivates performing local aggregation and replication over NVLink, reducing repeated network transfers, and initiating those transfers early. In a rail-optimized topology~\cite{wang2024railonly} commonly deployed today, it is more efficient to exchanges inter-node data between GPUs with
  \textit{the same local index}, while using NVLink for intra-node data movement.
  We refer to each such remote GPU as a \emph{rail peer}.

\paragraph{C2: Transfer-granularity asymmetry.}
Intra-node communication exposes peer-memory operations, whereas the network path uses explicit, message-oriented RDMA requests. Although RDMA addresses remote memory, each request specifies a buffer range and completion mechanism. In short, intra-node communication happens over \textit{memory semantics}; inter-node communication happens over \textit{message semantics}. Fused multi-node kernels must bridge these two interfaces: fine-grained local tiles should be grouped into network chunks to amortize per-message costs while exposing ready work early.

\paragraph{C3: Transport ordering semantics vary across platforms.}
An arrival flag must not become visible before its payload. InfiniBand reliable connections preserve the ordering of writes on a connection, whereas EFA's Scalable Reliable Datagram transport or the recent OpenAI MRC protocol~\cite{sohan2026mrc} does not provide the same guarantee~\cite{shalev2020srd,efadriver}. A shared kernel interface must therefore accommodate different completion mechanisms.

\paragraph{C4: Cross-device synchronization.}
Network completion and GPU memory visibility can also introduce performance overhead~\cite{gpudirectrdma}. A persistent kernel needs readiness checks and memory-ordering guarantees that let it consume remote data safely without global synchronization after each transfer. The challenge is to provide these guarantees with limited polling and synchronization overhead.

\paragraph{C5: Adapting SM allocation to the workload.}
The SM resources needed for intra- and inter-node communication vary across input shapes, workload mixes, and kernel types. Static sweeps can identify effective partitions~\cite{zhang2025comet,sul2025parallelkittens}, but repeated profiling and maintenance are needed as these choices change, especially as input shapes can be heterogenous. Our intra-node sweeps span best allocations of 2--64 communication SMs (\cref{fig:adaptive-sweep}). Additionally, the balance (the best number of SMs for communication versus compute) also changes within a single kernel, such as when there is work imbalance between compute and communication or the work cannot be neatly divided over the available SMs. Adaptive SM partition tuning must respond without excessive overhead.

\section{Design}
\label{sec:design}

\begin{figure*}[t]
\centering
\resizebox{\linewidth}{!}{%
\begin{tikzpicture}[x=1cm,y=1cm, font=\sffamily\small,
  region/.style={rounded corners=6pt, draw=none},
  role/.style={rounded corners=2pt, draw=none, text=white, font=\sffamily\footnotesize\bfseries, align=center, inner sep=0pt, minimum height=1.0cm},
  box/.style={rounded corners=2pt, draw=black!55, fill=white, align=center, font=\sffamily\footnotesize, inner sep=0pt},
  arr/.style={-{Stealth[length=2.2mm]}, line width=0.9pt, black!65},
  darr/.style={{Stealth[length=2.2mm]}-{Stealth[length=2.2mm]}, line width=0.9pt, black!65},
  num/.style={circle, fill=black!80, text=white, font=\sffamily\scriptsize\bfseries, inner sep=0pt, minimum size=4.2mm},
  rtitle/.style={anchor=west, font=\sffamily\footnotesize\bfseries},
  note/.style={font=\sffamily\scriptsize, text=mkgray, align=left}]
\fill[region, black!4] (0,0) rectangle (11.8,8.6);
\node[rtitle, font=\sffamily\small\bfseries] at (0.25,8.25) {Node 0};
\fill[region, mkorange!7] (0.3,3.5) rectangle (8.7,7.85);
\node[rtitle] at (0.5,7.5) {GPU 0: persistent kernel partitioned into SM roles};
\node[role, fill=mkblue,   minimum width=1.5cm] (send)  at (1.15,6.3) {Inter-node\\send};
\node[role, fill=mkviolet, minimum width=1.5cm] (recv)  at (2.8,6.3) {Inter-node\\receive};
\node[role, fill=mkorange, minimum width=1.5cm] (comp)  at (4.45,6.3) {Compute};
\node[role, fill=mkgray,   minimum width=1.5cm] (ctrl)  at (6.1,6.3) {Controller\\(SM split)};
\node[role, fill=mkaqua,   minimum width=1.5cm] (intra) at (7.75,6.3) {Intra-node\\comm.};
\node[box, minimum width=6.45cm, minimum height=0.8cm] (mem) at (5.275,4.4) {GPU memory: buffers, ready flags, progress counters};
\draw[arr]  (2.8,4.8) -- (2.8,5.8);
\draw[arr]  (4.45,5.8) -- (4.45,4.8);
\draw[arr]  (6.1,4.8) -- (6.1,5.8);
\draw[darr] (7.75,5.8) -- (7.75,4.8);
\draw[line width=0.8pt, black!65, dashed] (6.1,6.8) -- (6.1,7.05) -- (1.15,7.05);
\draw[line width=0.8pt, black!65, dashed] (6.1,7.05) -- (7.75,7.05);
\foreach \x in {1.15,2.8,4.45,7.75} { \draw[arr, dashed] (\x,7.05) -- (\x,6.82); }
\node[note, anchor=south] at (3.6,7.06) {target split};
\node[box, fill=mkaqua!15, minimum width=1.6cm, minimum height=1.0cm] (nvs) at (10.0,6.3) {NVSwitch};
\node[box, minimum width=1.6cm, minimum height=0.7cm] (peers) at (10.0,7.6) {GPUs 1--7};
\draw[darr] (8.5,6.3) -- (9.2,6.3);
\draw[darr] (10.0,6.8) -- (10.0,7.25);
\fill[region, black!9] (0.3,0.3) rectangle (6.6,2.7);
\node[anchor=east, font=\sffamily\footnotesize\bfseries] at (6.45,2.4) {Host CPU};
\node[box, minimum width=2.35cm, minimum height=1.2cm] (queue) at (1.725,1.3) {Command\\queue};
\node[box, minimum width=2.9cm,  minimum height=1.2cm] (proxy) at (4.75,1.3) {Proxy thread};
\draw[arr] (1.15,5.8) -- (1.15,1.9);
\draw[arr] (2.9,1.3) -- (3.3,1.3);
\node[box, fill=mkblue!12, minimum width=2.2cm, minimum height=1.2cm] (nic) at (9.5,1.3) {RDMA NIC};
\draw[arr] (6.2,1.3) -- (8.4,1.3);
\draw[arr] (8.5,4.4) -- (10.1,4.4) -- (10.1,1.9);
\node[note, anchor=west] at (10.25,3.15) {GPUDirect\\RDMA read};
\fill[region, black!4] (13.0,0) rectangle (18.6,8.6);
\node[rtitle, font=\sffamily\small\bfseries] at (13.25,8.25) {Node 1};
\fill[region, mkorange!7] (13.3,3.5) rectangle (18.3,7.85);
\node[rtitle] at (13.5,7.5) {GPU 0 (rail peer)};
\node[role, fill=mkviolet, minimum width=2.1cm] (recv1) at (14.65,6.3) {Inter-node\\receive};
\node[role, fill=mkorange, minimum width=2.1cm] (comp1) at (16.95,6.3) {Compute};
\node[box, minimum width=4.4cm, minimum height=0.8cm] (mem1) at (15.8,4.4) {receive buffer, arrival flags};
\draw[arr] (14.3,4.8) -- (14.3,5.8);
\node[box, fill=mkblue!12, minimum width=2.2cm, minimum height=1.2cm] (nic1) at (15.2,1.3) {RDMA NIC};
\draw[arr] (15.2,1.9) -- (15.2,4.0);
\draw[arr, line width=1.2pt, mkblue!85!black] (10.6,1.3) -- (14.1,1.3);
\node[note, align=center, anchor=north] at (12.4,1.05) {RDMA write\\and arrival flag};
\node[num] at (4.85,5.3) {1};
\node[num] at (8.85,6.72) {2};
\node[num] at (1.55,3.1) {3};
\node[num] at (7.3,1.7) {4};
\node[num] at (12.4,1.7) {5};
\node[num] at (14.68,5.3) {6};
\end{tikzpicture}}
\caption{\textbf{Architecture of \system} (two nodes; one GPU per node shown in detail). Each GPU executes a persistent kernel whose thread blocks are assigned to roles. \textcircled{\scriptsize 1} Compute blocks write completed tiles to GPU memory and set readiness flags. \textcircled{\scriptsize 2} Intra-node communication blocks exchange tiles with the other seven GPUs over NVLink, using NVSwitch for broadcast and reduction. \textcircled{\scriptsize 3} Inter-node send blocks enqueue transfer commands in host memory. \textcircled{\scriptsize 4} A host proxy thread submits batches of commands to the NIC. \textcircled{\scriptsize 5} The NIC transfers data to the rail peer on node 1; the NIC or receiving proxy signals arrival, depending on the transport. \textcircled{\scriptsize 6} Receive blocks observe the flag and consume the data. The optional controller (\cref{sec:adaptive}) reads progress counters and publishes a target number of communication blocks (dashed).}%
\label{fig:overview}
\end{figure*}
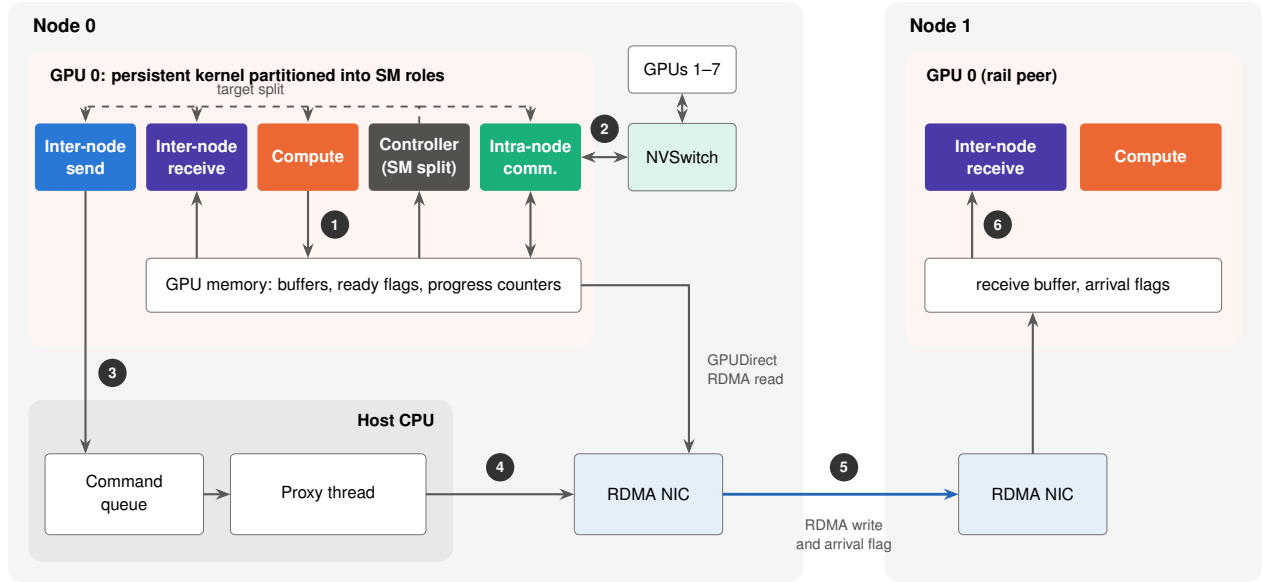

\Cref{fig:overview} shows how \system overlaps computation, NVLink transfers, and inter-node RDMA inside persistent kernels. The central design problem is to preserve tile-level progress across both intra- and inter- node communication boundary: a locally ready tile is not necessarily sent out as a complete network message, and a submitted message is not yet safe for remote computation to consume. \system separates these stages through explicit readiness handoffs. We use ThunderKittens' compute primitives~\cite{spector2025thunderkittens}; this section focuses on coordinating them with intra- and inter-node communication.

Thread blocks specialize in computation, intra-node communication, inter-node submission, or receive-side processing. Dedicated thread blocks submit ready work while compute blocks continue producing tiles. A host proxy forwards commands to the NIC; payloads remain in GPU memory. The following subsections cover hierarchical scheduling and transfer granularity (\cref{sec:intra}), adaptive SM allocation (\cref{sec:exec}), and inter-node batching and completion (\cref{sec:inter}).

\subsection{Bridging intra- and inter-node communication granularity}
\label{sec:intra}

\paragraph{Reduce or broadcast locally via NVSwitch before exchanging data across nodes.}
  \system uses NVSwitch for local reduction and broadcast to minimize redundant inter-node
traffic (C1). For example, on GEMM+AllReduce, each GPU in a node is responsible for one eighth of the
output tiles. NVSwitch reduces the eight local GPUs' contributions to each tile. The
responsible GPU then exchanges this partial sum with its rail peer (GPUs with the same GPU index across nodes), combines the local and
remote sums, and broadcasts the completed tile within its node. These steps proceed
independently for each tile. For AllGather+GEMM, each shard is sent once per destination
node, where the receiving rail peer broadcasts it to the other local GPUs over NVLink.

\paragraph{Differentiate tile readiness and message readiness.}
\label{sec:granularity}
Within a node, communication uses peer-memory operations on tiles; across nodes, RDMA requires explicit requests specifying buffer ranges and completion information. Mapping every local tile to a network request would tie local parallelism to per-message overhead (C2). \system instead sizes compute tiles, local transfers, and network chunks independently. Larger network chunks amortize submission and notification costs, but wait for more data and can lengthen pipeline fill and drain.

For example, GEMM+AllReduce groups four locally reduced tiles into a 256\,KiB network chunk. Local thread blocks process tiles independently; a per-chunk counter lets the last completed tile publish readiness to the sender. A chunk can then be sent out (either with intra-node communication or inter-node communication) without waiting for the remaining GEMM computation to finish. The inverse mapping matters at the receiver. For example, Dispatch+GEMM checks every 512\,KiB network chunk intersecting a token's byte range before loading that token, A token spanning a chunk boundary requires both remote message arrivals. Since there is data dependency between computation and communication, in our case, either communication supplies input to computation, or vice versa.

\begin{takeaway}
  \textbf{Takeaway 1.} Hierarchical communication reduces traffic on the slower inter-node network, while
  independently sizing local tiles and network chunks balances early transmission against per-message
  overhead. \system uses NVSwitch for local aggregation and replication and sends each RDMA chunk as soon as
  its constituent tiles are ready.
  \end{takeaway}

\subsection{SM partioning and adaptive tuning}
\label{sec:exec}

The preceding mechanisms determine when work is ready; the SM partitioning determines how quickly each stage can progress. Thread blocks are assigned roles by block index in the fixed configuration and may change roles at task boundaries in adaptive mode. Separate SM roles let \system change the allocation without changing the compute block's warp layout, building on the scheduling choices.

\begin{figure*}[t]
\centering
\includegraphics[width=.8\linewidth]{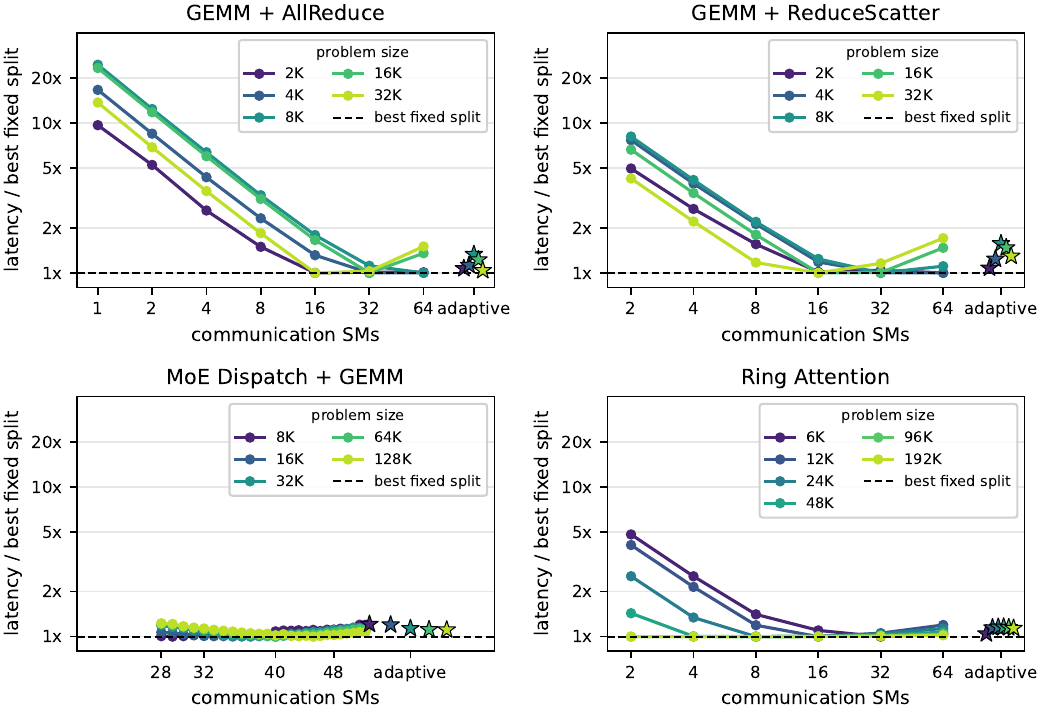}
\caption{\textbf{Static sweep of the SM partition} (single node, 8 H100 GPUs). Each curve gives the latency of one problem size as a function of the number of communication SMs, normalized to the best fixed SM partition for that size (dashed line). The star markers at the right of each panel give the latency of the adaptive controller for the same sizes.}
\label{fig:adaptive-sweep}
\end{figure*}

\paragraph{Adaptive tuning of the SM partition.}
\label{sec:adaptive}
Static sweeps identify efficient SM partitions, but require choices to be profiled and maintained for each workload (C5).
\system provides an adaptive mode that updates the allocation within an execution. Each block checks a shared \emph{target} number of communication blocks between compute or communication tasks (\textit{e.g.},  a tile or a message) and changes roles when the active allocation differs from this target.
\Cref{fig:adaptive-sweep} compares this policy with static sweeps for four intra-node kernels.

The controller estimates the per-task cost of each role from cycle counters accumulated by the thread blocks and sets the target to the value that equalizes the predicted completion times of the two roles,
\begin{equation}
n_s^{\ast} \;=\; B\cdot\frac{R_s\,C_s}{R_p\,C_p + R_s\,C_s},
\label{eq:adaptive}
\end{equation}
where $R_p$ and $R_s$ are the remaining compute and communication tasks, and $C_p$ and $C_s$ are their estimated costs.

The data dependency determines the initial allocation and how stalled work is treated. When communication supplies inputs to computation, a compute block waiting for input may temporarily assist communication even if the target has been reached.
We evaluate adaptive tuning on intra-node kernels in \cref{sec:eval-adaptive}.

\begin{takeaway}
\textbf{Takeaway 2.} SM specialization separates role implementation from resource allocation. \system can rebalance compute, intra-node communication, and inter-node communication by adjusting their SM budgets without redesigning the compute block's warp layout.
\end{takeaway}

\subsection{Portable inter-node communication in fused kernels}
\label{sec:inter}

The host-assisted GPU-initiated communication path translates GPU-produced work into explicit RDMA messages. A proxy implemented on \texttt{libibverbs} handles submission and transport-specific notification, following the GPU-command/CPU-proxy design in UEP~\cite{mao2025ucclep}.

\paragraph{Command publication and backpressure.}
Send blocks publish 48-byte commands in a ring buffer in pinned host memory. Each command specifies the peer, source and destination offsets, byte count, and chunk identifier. The GPU writes the command body before committing its header; the proxy polls this header before reading the record. Queue credits and a limit on outstanding requests bound the work submitted to the host and NIC. The NIC reads the payload directly from registered GPU memory, including staging buffers when the network layout differs from the compute layout.

\paragraph{Batch commands while preserving per-chunk completion.}
The proxy groups up to eight ready commands for one connection into a single submission. This is distinct from constructing a larger network chunk: each command retains its own data transfer and arrival notification. A fixed requirement to fill every batch would delay sparse arrivals and the final few chunks. Instead, the proxy submits partial batches, for example, using a bounded polling window to collect additional transfer commands. Submission batching thus follows the rate at which the kernel produces ready work, while chunk size controls when that work first becomes transferable.

\paragraph{Transport delivery semantics vary across platforms.}
An arrival signal must identify a completed payload, not merely a posted request (C3). InfiniBand and EFA require different notification paths (\cref{tab:backends}):
\begin{itemize}
  \item \emph{InfiniBand (ConnectX-7).} The proxy posts a data write followed by a small flag write on the same reliable connection (RC). Receive blocks poll the flag in GPU memory, without a receiving host proxy. This protocol requires data-before-flag ordering at the destination.
  \item \emph{AWS EFA (SRD).} SRD does not order independent writes~\cite{efadriver}. The completion-based path therefore uses RDMA write with immediate data: the immediate value identifies the chunk, and the receiving proxy publishes its arrival flag after the write's receive completion. This avoids using the arrival order of a separate flag write to infer payload completion.
\end{itemize}

\begin{table}[t]
\centering
\small
\caption{Inter-node notification paths (EFA's completion-based variant shown). Transport completion and GPU memory visibility are separate requirements.}
\label{tab:backends}
\begin{tabular}{@{}l l l@{}}
\toprule
 & \textbf{ConnectX-7 (InfiniBand)} & \textbf{AWS EFA (SRD)} \\
\midrule
Transport order      & ordered per RC connection & unordered \\
Notification path    & data write, then flag write & write with immediate; proxy sets flag \\
\bottomrule
\end{tabular}
\end{table}

\paragraph{Host-assisted GPU-initiated communication versus GPUDirect Async communication.}
\label{sec:design-ibgda}

\begin{figure}[t]
\centering
\includegraphics[width=0.78\linewidth]{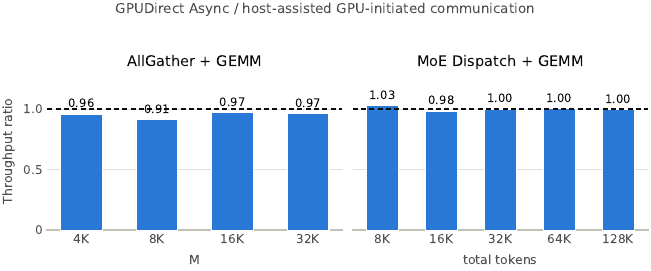}
\caption{\textbf{GPUDirect Async (IBGDA) relative to host-assisted GPU-initiated communication} on the ConnectX-7 testbed of \cref{sec:eval-setup} (2 nodes $\times$ 8 H200), with identical kernels. Bars give the throughput ratio, with GPUDirect Async in the numerator; the dashed line marks parity.}
\label{fig:ibgda}
\end{figure}

We also implement GPUDirect Async on ConnectX-7 using IBGDA~\cite{ibgda}, exposing the same device interface as host-assisted GPU-initiated communication. \Cref{fig:ibgda} compares the two backends. In the scenarios we tested, GPUDirect Async provides little performance gain over host-assisted GPU-initiated communication.

Host-assisted GPU-initiated communication lets the proxy overlap submission with computation and batch multiple commands, while the GPUDirect Async backend needs to issue an \textit{expensive} system-scoped GPU fence and doorbell per chunk.
GPUDirect Async (IBGDA) can also be used in small-message, latency-sensitive workloads~\cite{ibgda,deepep2025}.

\begin{takeaway}
\textbf{Takeaway 3.} Local tiles, network chunks, and submission batches need not share one granularity. A batched host proxy connects these stages while keeping payload transfers on the NIC. GPUDirect Async (IBGDA) provides no consistent throughput advantage over host-assisted GPU-initiated communication in our comparison.
\end{takeaway}

\section{Implementation}
\label{sec:kernels}

\Cref{tab:ops} summarizes the five kernels implemented in \system. They use the mechanisms in \cref{sec:design}: persistent scheduling with explicit SM roles, hierarchical communication through rail peers, and epoch-stamped readiness flags. \system makes three scheduling choices: First, inter-node transfers are initiated early: AllGather+GEMM and Ring Attention submit their input shard or KV slice to rail peers at launch. Second, NVSwitch handles local aggregation and replication. For example, GEMM+AllReduce reduces data within NVSwitch before each GPU sends one eighth of the output. Third, compute tiles are ordered according to input availability: AllGather+GEMM processes its local shard, the remaining shards within the node, and then remote shards.

\begin{table*}[t]
\centering
\footnotesize
\setlength{\tabcolsep}{4pt}
\renewcommand{\arraystretch}{1.2}
\caption{Fused kernels implemented in \system. TP, SP, and EP denote tensor, sequence, and expert parallelism.
}
\label{tab:ops}
\begin{tabular}{@{}>{\raggedright\arraybackslash}p{2.8cm} >{\raggedright\arraybackslash}p{2.0cm} >{\raggedright\arraybackslash}p{3.6cm} >{\raggedright\arraybackslash}p{4.2cm} >{\raggedright\arraybackslash}p{2.3cm}@{}}
\toprule
\textbf{Kernel} & \textbf{Data dependency} & \textbf{Intra-node} & \textbf{Inter-node} & \textbf{Chunk} \\
\midrule
AllGather + GEMM (TP) & comm.\ $\rightarrow$ compute & multicast broadcast of each shard & shard sent once to each node; ring forwarding beyond two nodes & 128 rows \\
GEMM + ReduceScatter (TP) & compute $\rightarrow$ comm. & TMA atomic add into the owner's buffer & exchange of per-node partial sums & 2--32 tiles \\
GEMM + AllReduce (TP) & compute $\rightarrow$ comm. & NVSwitch reduction; multicast broadcast of the result & exchange of 1/8 of the output per GPU & 4 tiles (256\,KiB) \\
MoE Dispatch + GEMM (EP) & comm.\ $\rightarrow$ compute & TMA pull of tokens from peers & token buffer copied to the rail peer & 16 tokens; 512\,KB \\
Ring Attention (SP) & independent within a step & TMA store of KV to the next GPU & KV slice sent once to each node & 128-row KV tiles \\
\bottomrule
\end{tabular}
\end{table*}

\section{Evaluation}
\label{sec:eval}

Our evaluation addresses four questions:
\begin{enumerate}
  \item How do the fused kernels perform relative to unfused execution such as using cuBLAS or FlashAttention and NCCL (\cref{sec:eval-tp,sec:eval-ep,sec:eval-sp})?
  \item How does \system compare with existing fused kernels (\cref{sec:eval-tp,sec:eval-ep,sec:eval-sp})?
  \item How sensitive is performance to the compute--communication SM partition, and how close does the adaptive controller come to the best fixed SM partition (\cref{sec:eval-adaptive})?
\end{enumerate}

\subsection{Experimental setup}
\label{sec:eval-setup}

\begin{table}[t]
\centering
\small
\caption{Testbeds. Both contain 2 nodes $\times$ 8 H200 GPUs connected by NVLink/NVSwitch within each node, with different inter-node networks.}
\label{tab:testbeds}
\begin{tabular}{@{}l l l l@{}}
\toprule
\textbf{Testbed} & \textbf{GPUs} & \textbf{Inter-node network} & \textbf{NICs per node} \\
\midrule
ConnectX-7 & 2 $\times$ 8 H200 & InfiniBand             & 8 $\times$ 400\,Gb/s ConnectX-7 \\
AWS EFA    & 2 $\times$ 8 H200 & AWS SRD                & 16 $\times$ 200\,Gb/s EFA \\
\bottomrule
\end{tabular}
\end{table}

\paragraph{Testbeds.}
\Cref{tab:testbeds} describes the two clusters, both of which provide 400\,Gb/s of network bandwidth per GPU. \system is compiled for Hopper (\texttt{sm\_90a}) with CUDA 12.9, and one process runs per GPU.

\paragraph{Baselines.}
\emph{unfused} execution: a cuBLAS GEMM or FlashAttention~\cite{dao2023flashattention} kernel preceded or followed by the corresponding NCCL collective (all-gather, reduce-scatter, all-reduce, or all-to-all). We also compare against the following fused/comm-comp overlapped kernels/systems:
\begin{itemize}
  \item Triton-distributed~\cite{zheng2025tritondistributed} and FLUX~\cite{chang2024flux}, which fuse GEMMs with communication;
  \item Mercury~\cite{guan2025mercury}, a multi-GPU kernel compiler;
  \item DeepEP~\cite{deepep2025} combined with DeepGEMM~\cite{deepgemm2025}, for MoE dispatch;
  \item MagiAttention~\cite{magiattention} and ring-flash-attention~\cite{ringflashattn}, for distributed attention.
\end{itemize}
Each baseline is evaluated on the kernels and testbeds supported by the implementation used in the experiments. The evaluated DeepEP implementation uses GPUDirect Async (IBGDA) on InfiniBand and does not support EFA. The EFA dispatch plot therefore includes its ConnectX-7 measurements as a cross-testbed reference.

Measurements follow a warm-up phase and use the slowest rank's execution time, which determines when the next layer can begin. Kernels are timed over consecutive launches with one synchronization at the end.

\begin{figure*}[p]
\centering
\includegraphics[width=0.49\linewidth]{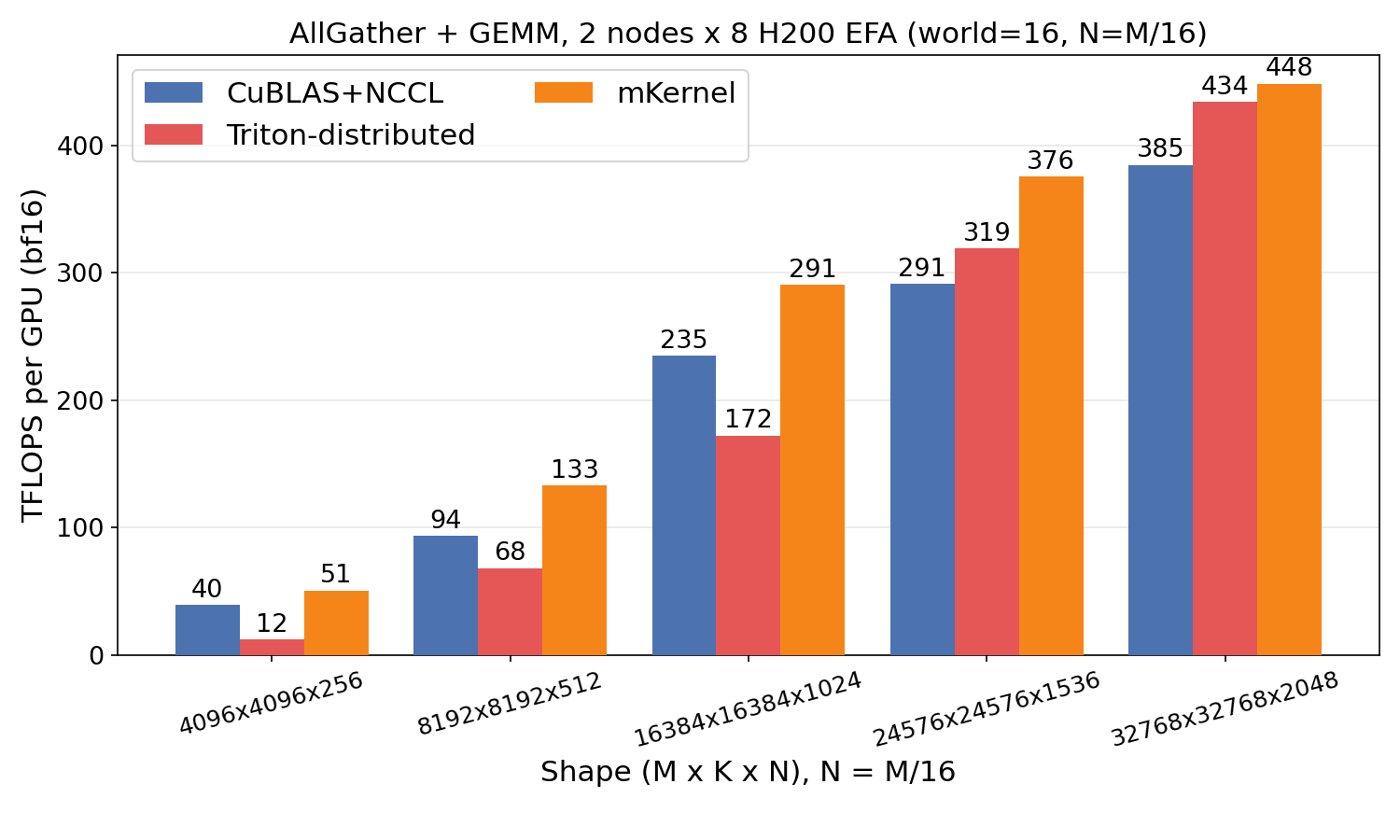}\hfill
\includegraphics[width=0.49\linewidth]{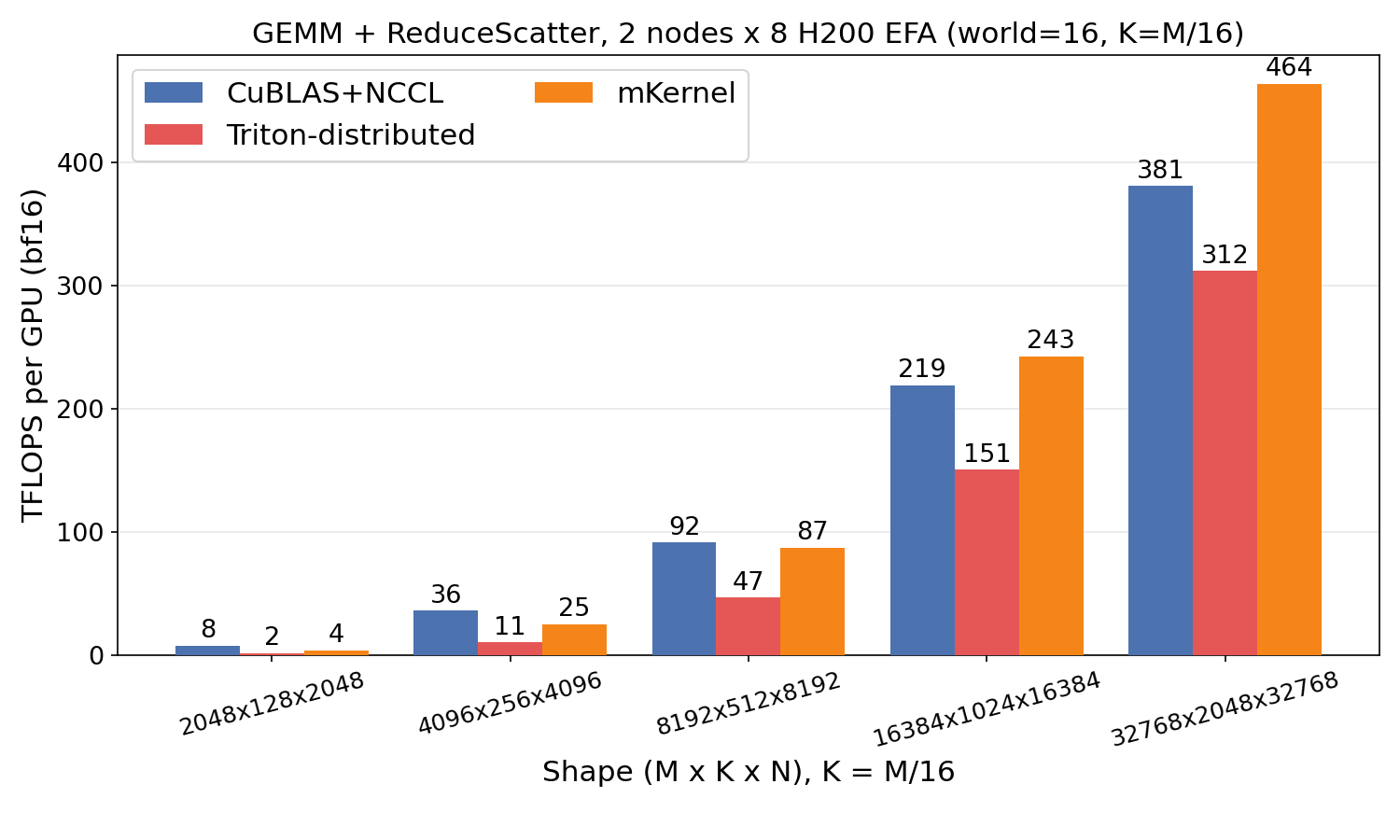}\\[4pt]
\includegraphics[width=0.49\linewidth]{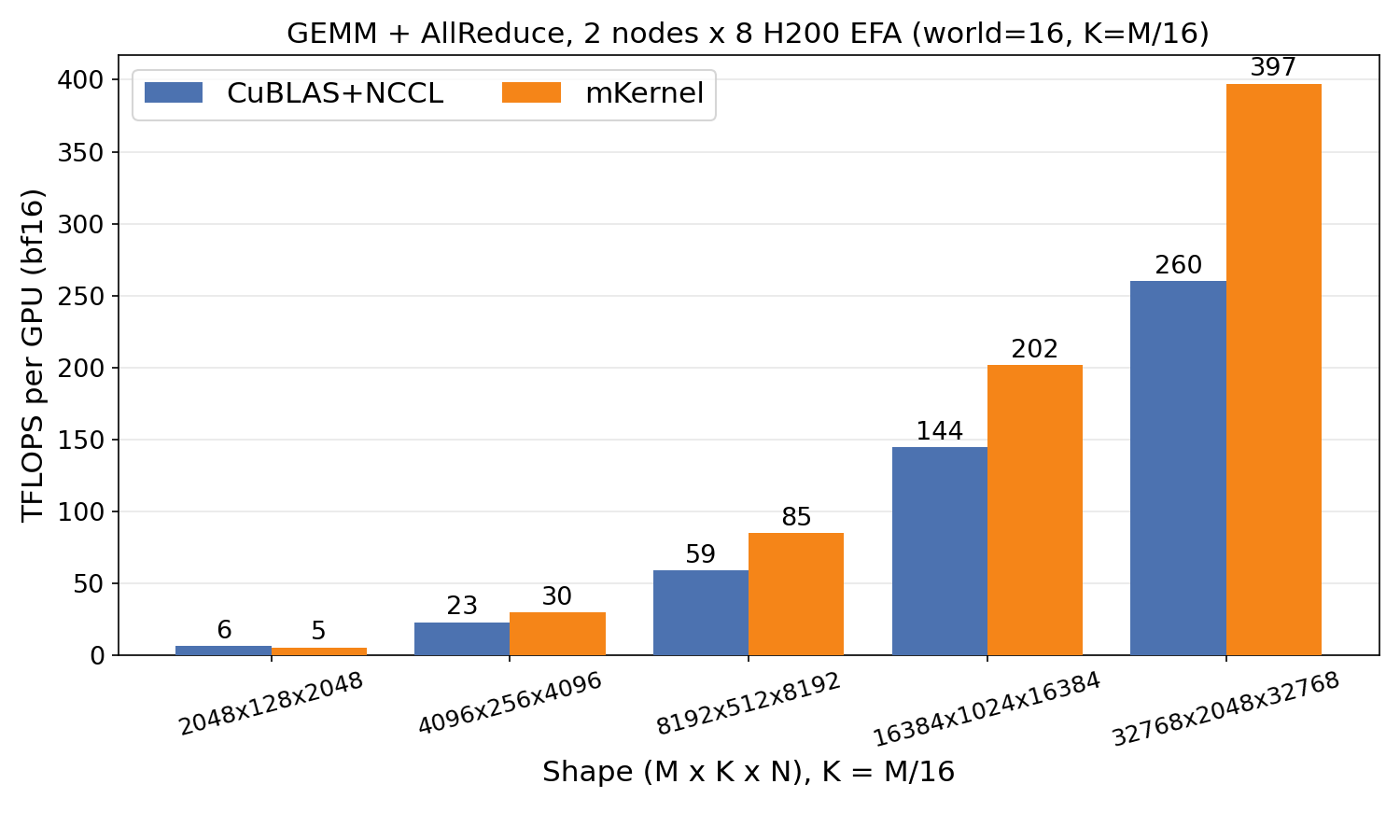}\hfill
\includegraphics[width=0.49\linewidth]{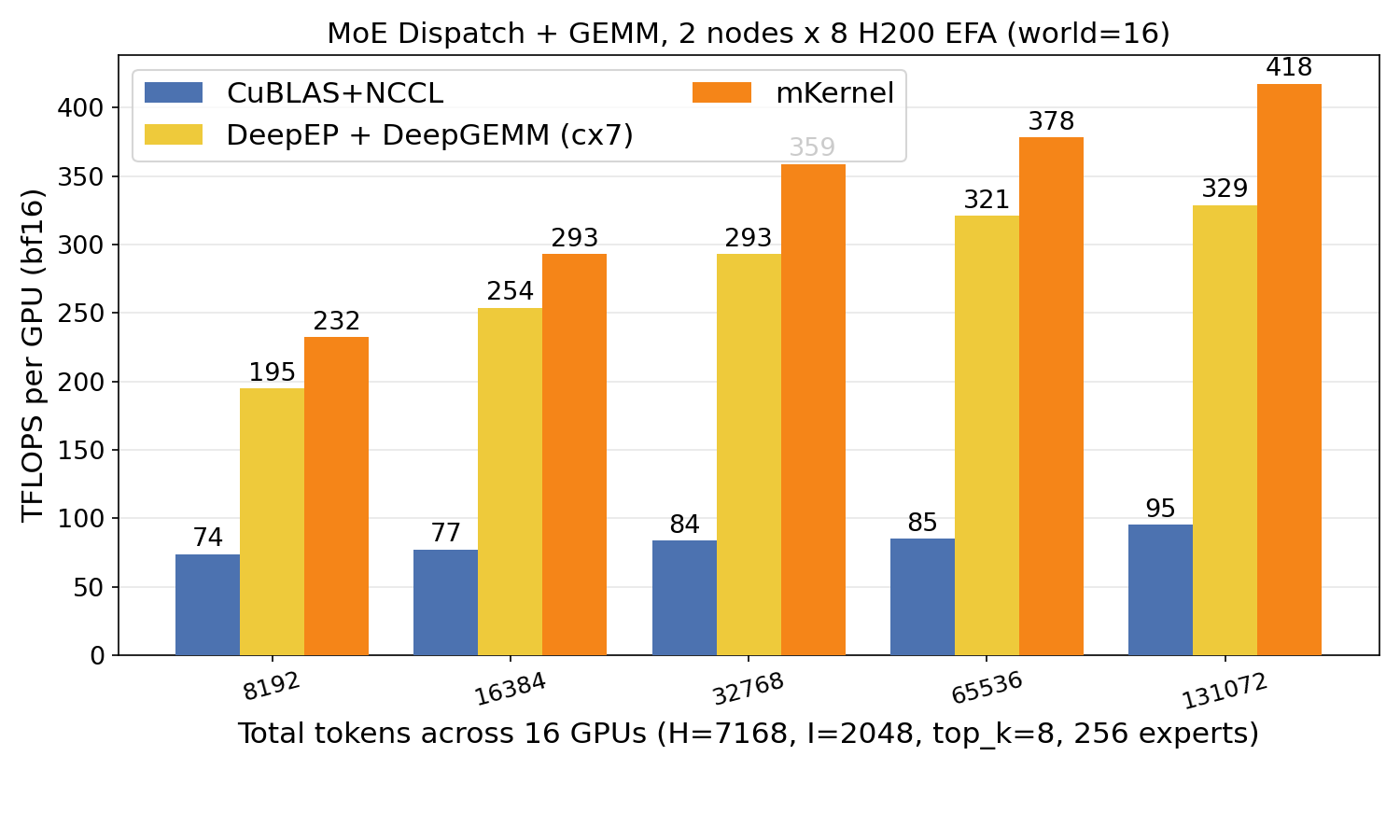}\\[4pt]
\includegraphics[width=0.49\linewidth]{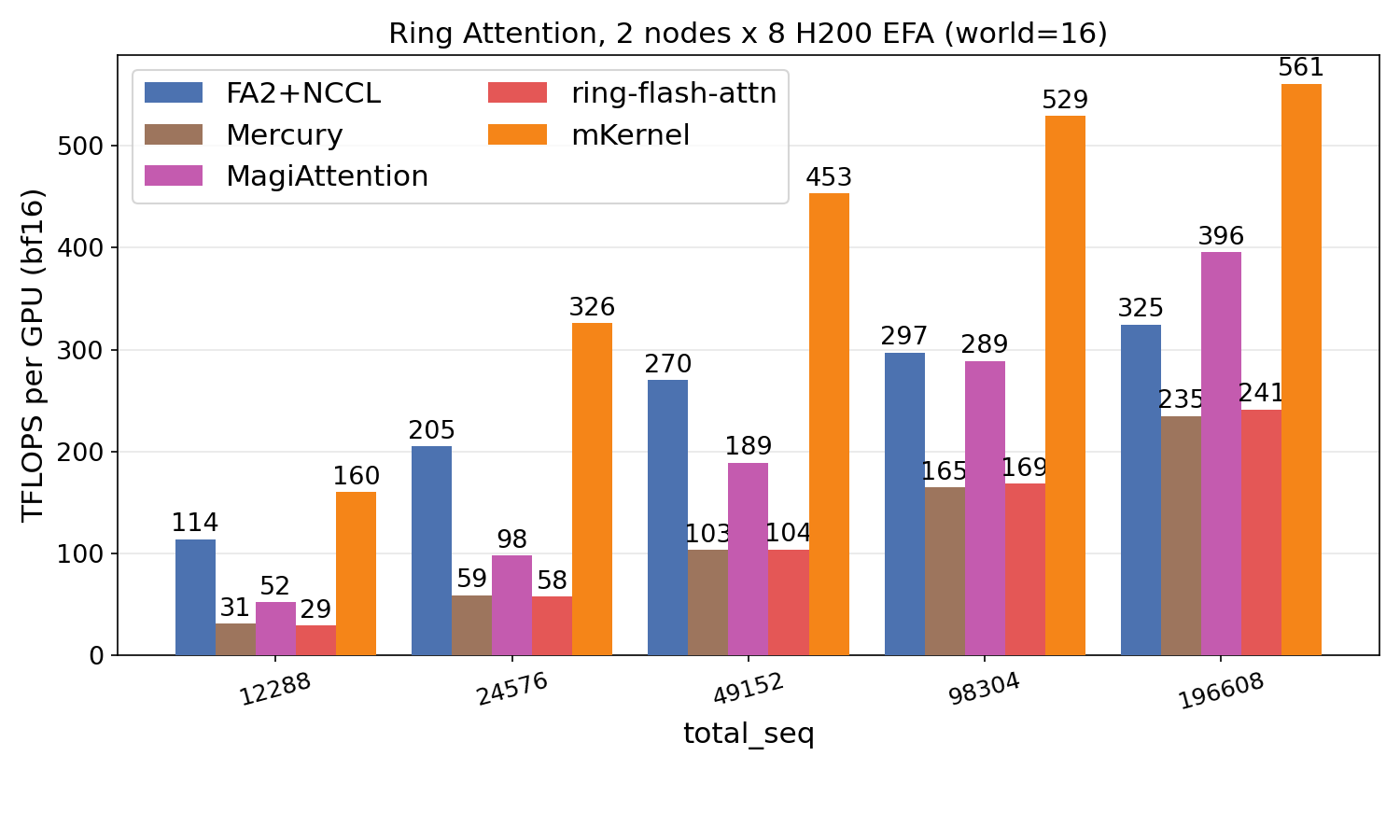}
\caption{\textbf{Throughput on the AWS EFA testbed} (2 nodes $\times$ 8 H200; TFLOPS per GPU, higher is better). \system is shown in orange. The evaluated DeepEP+DeepGEMM implementation does not support EFA; its bars show ConnectX-7 measurements for reference.}
\label{fig:eval-efa}
\end{figure*}

\begin{figure*}[tbp]
\centering
\includegraphics[width=0.49\linewidth]{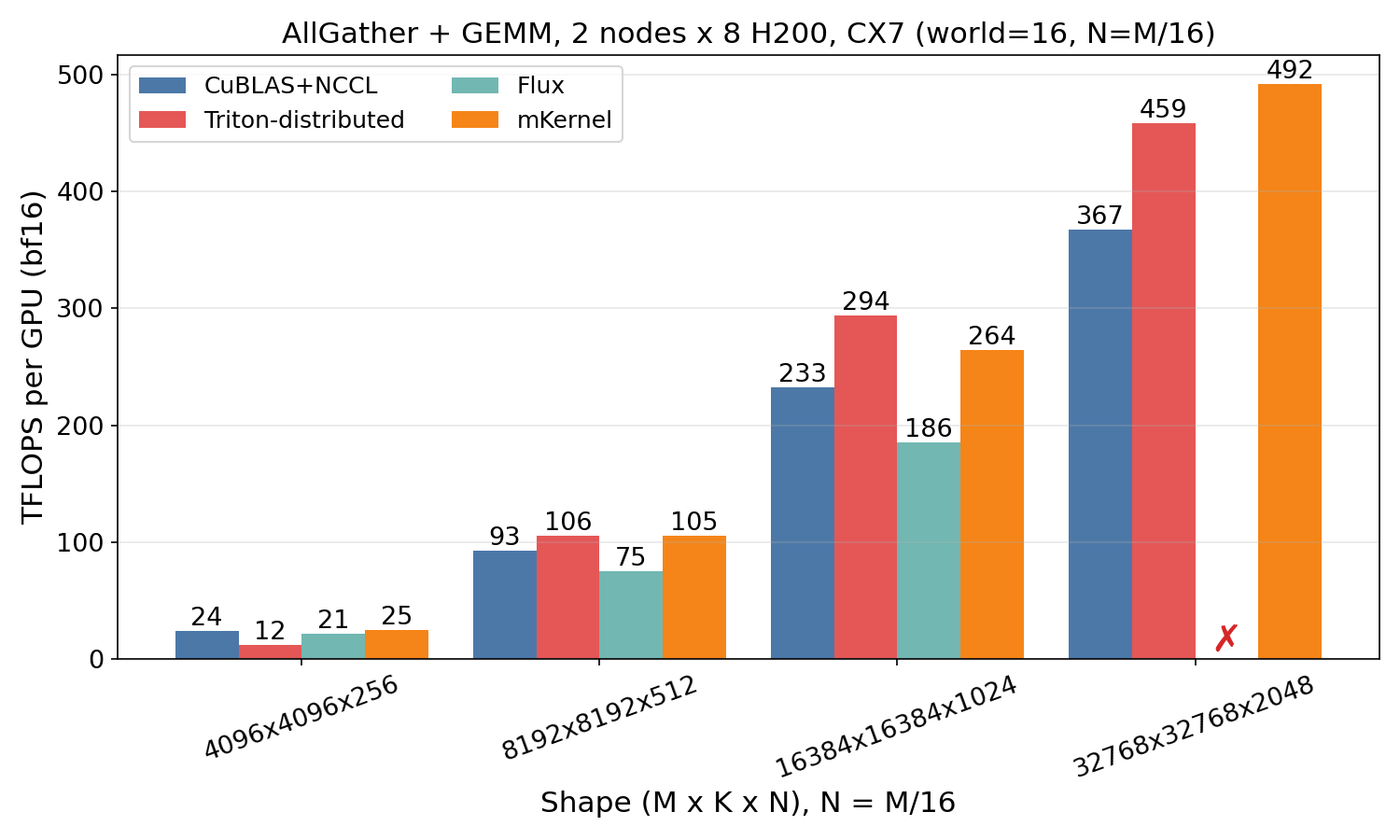}\hfill
\includegraphics[width=0.49\linewidth]{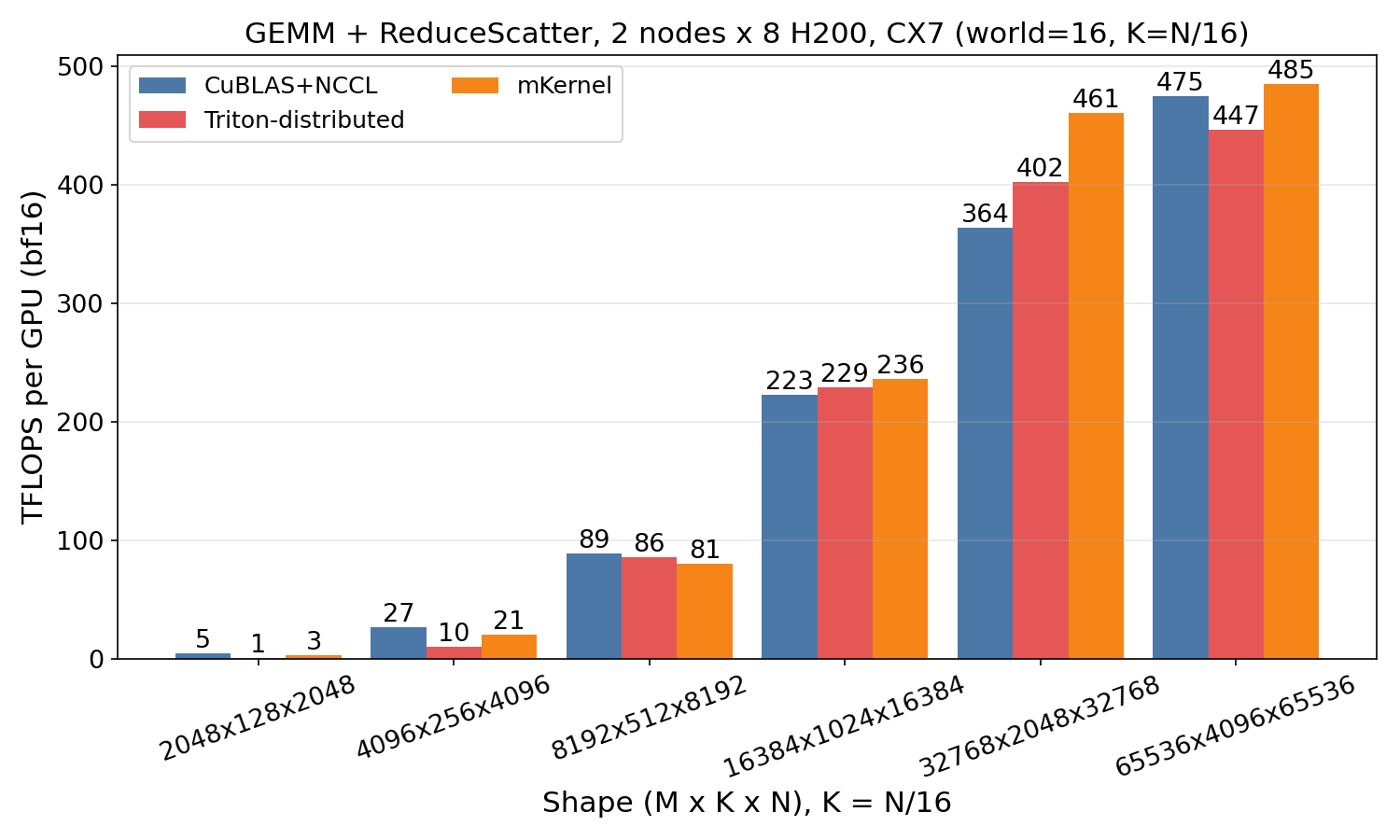}\\[4pt]
\includegraphics[width=0.49\linewidth]{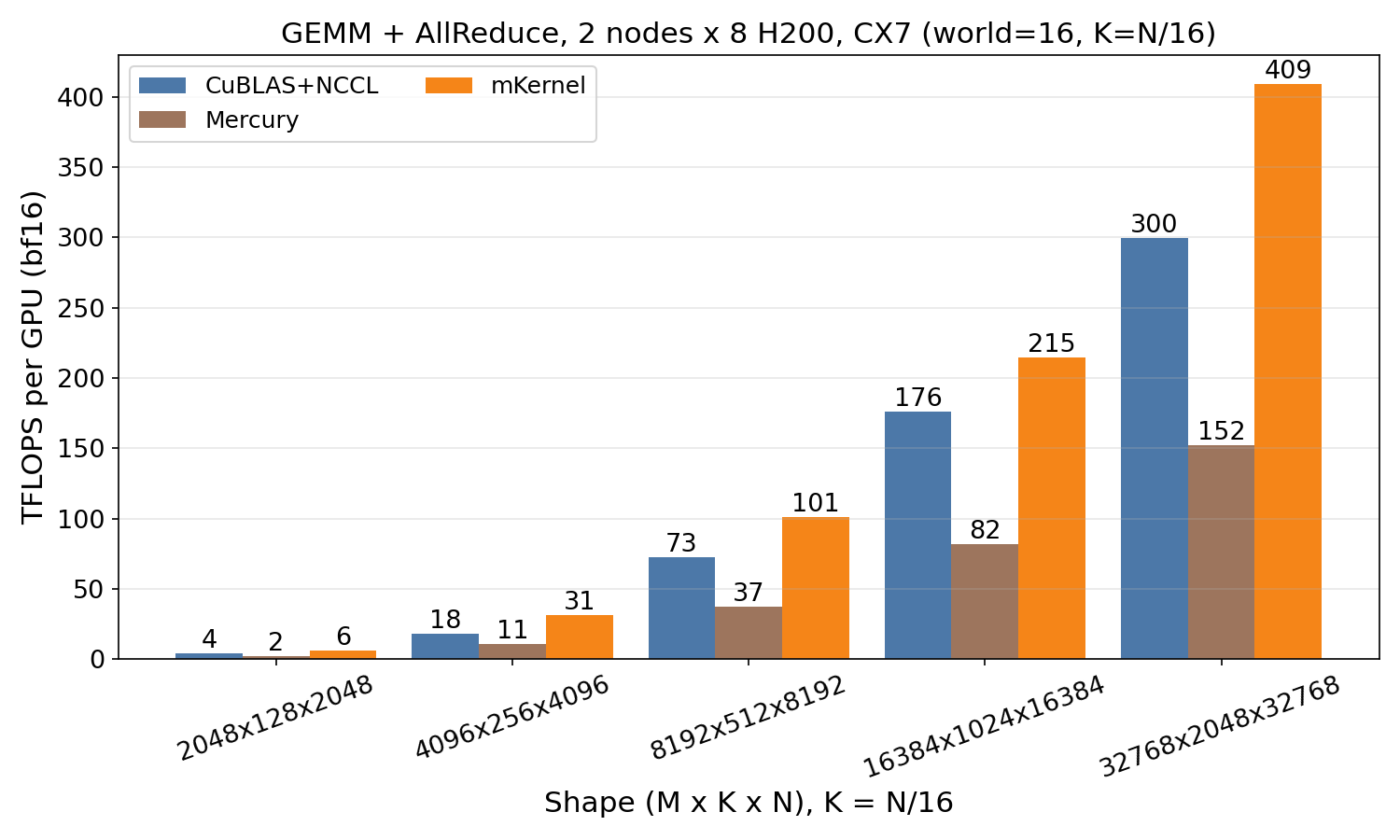}\hfill
\includegraphics[width=0.49\linewidth]{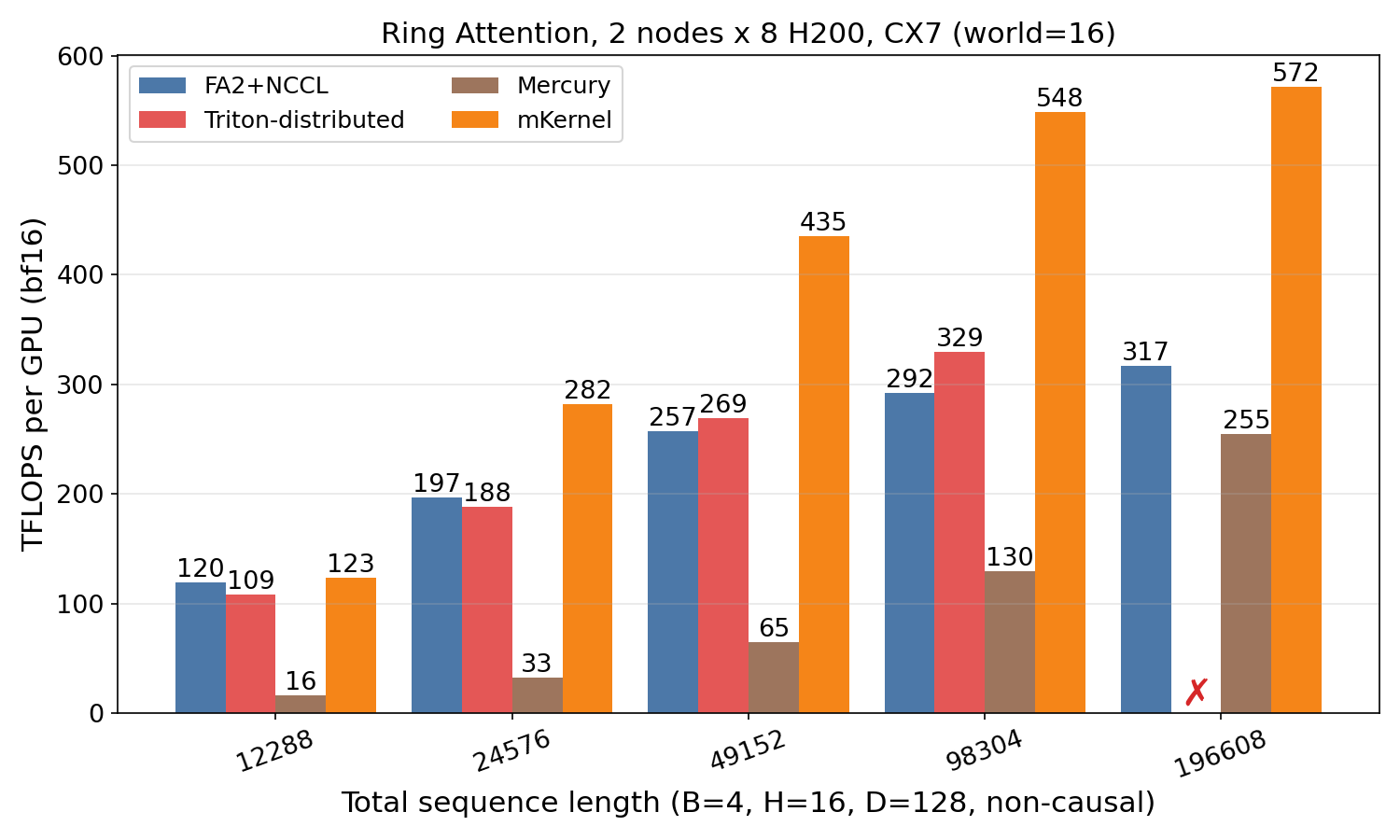}
\caption{\textbf{Throughput on the ConnectX-7 InfiniBand testbed} (2 nodes $\times$ 8 H200; TFLOPS per GPU). A red $\times$ indicates that a baseline failed at that problem size.}
\label{fig:eval-cx7}
\end{figure*}

\Cref{fig:eval-efa,fig:eval-cx7} present the results on EFA and ConnectX-7, respectively.

\subsection{Tensor parallelism}
\label{sec:eval-tp}

\paragraph{AllGather + GEMM.}
\system outperforms cuBLAS+NCCL at every evaluated problem size on both testbeds, with speedups of up to $1.41\times$ on EFA and $1.34\times$ on ConnectX-7.
The schedule initiates inter-node transfers first and transfers each remote shard once per destination node, allowing communication to overlap with GEMM.

\paragraph{GEMM + AllReduce.}
GEMM+AllReduce has the largest peak speedup among the evaluated tensor-parallel kernels. \system achieves up to $1.53\times$ on EFA and $1.72\times$ on ConnectX-7.
The gain is due to \system's  hierarchical schedule: intra-node reductions are performed in NVSwitch; each GPU sends only 1/8 of the output across the network; all three phases proceed concurrently with the GEMM.

\paragraph{GEMM + ReduceScatter.}
\system is faster for large problems ($1.02$--$1.27\times$ for $M \ge 16$K) and outperforms Triton-distributed at most input size.

\subsection{Expert parallelism}
\label{sec:eval-ep}

\paragraph{MoE Dispatch + GEMM.}
On EFA, \system
achieves speedups of $3.1$--$4.5\times$ over an all-to-all followed by a GEMM. This baseline assumes uniform routing and does not group expert GEMMs, so the comparison includes benefits from expert computation organization as well as communication overlap. Our kernel also outperforms DeepEP+DeepGEMM (EFA); where the DeepEP+DeepGEMM bars are measurements on ConnectX-7, due to testbed limitations.

\subsection{Sequence parallelism}
\label{sec:eval-sp}
\system's Ring Attention kernel achieves speedups of up to $1.78\times$ on EFA and $1.88\times$ on ConnectX-7 over the unfused baseline.
Against the evaluated distributed-attention implementations, \system achieves the following speedups:
\begin{itemize}
  \item $1.4$--$3.3\times$ over MagiAttention;
  \item $2.3$--$5.6\times$ over ring-flash-attention;
  \item $2.2$--$8.6\times$ over Mercury;
  \item $1.1$--$1.7\times$ over Triton-distributed, whose evaluated implementation fails on the longest sequence.
\end{itemize}
The advantage is generally larger for shorter sequences, where communication is harder to amortize over attention computation.

\subsection{Adaptive SM partitioning}
\label{sec:eval-adaptive}

A static sweep measures sensitivity to the SM partition and identifies the best measured configuration for each problem size. We then compare the adaptive controller in \cref{sec:adaptive} with those configurations. Due to testbed limitations, we are only able to test adaptive SM partitioning on a single node.

\begin{figure}[t]
\centering
\includegraphics[width=.8\linewidth]{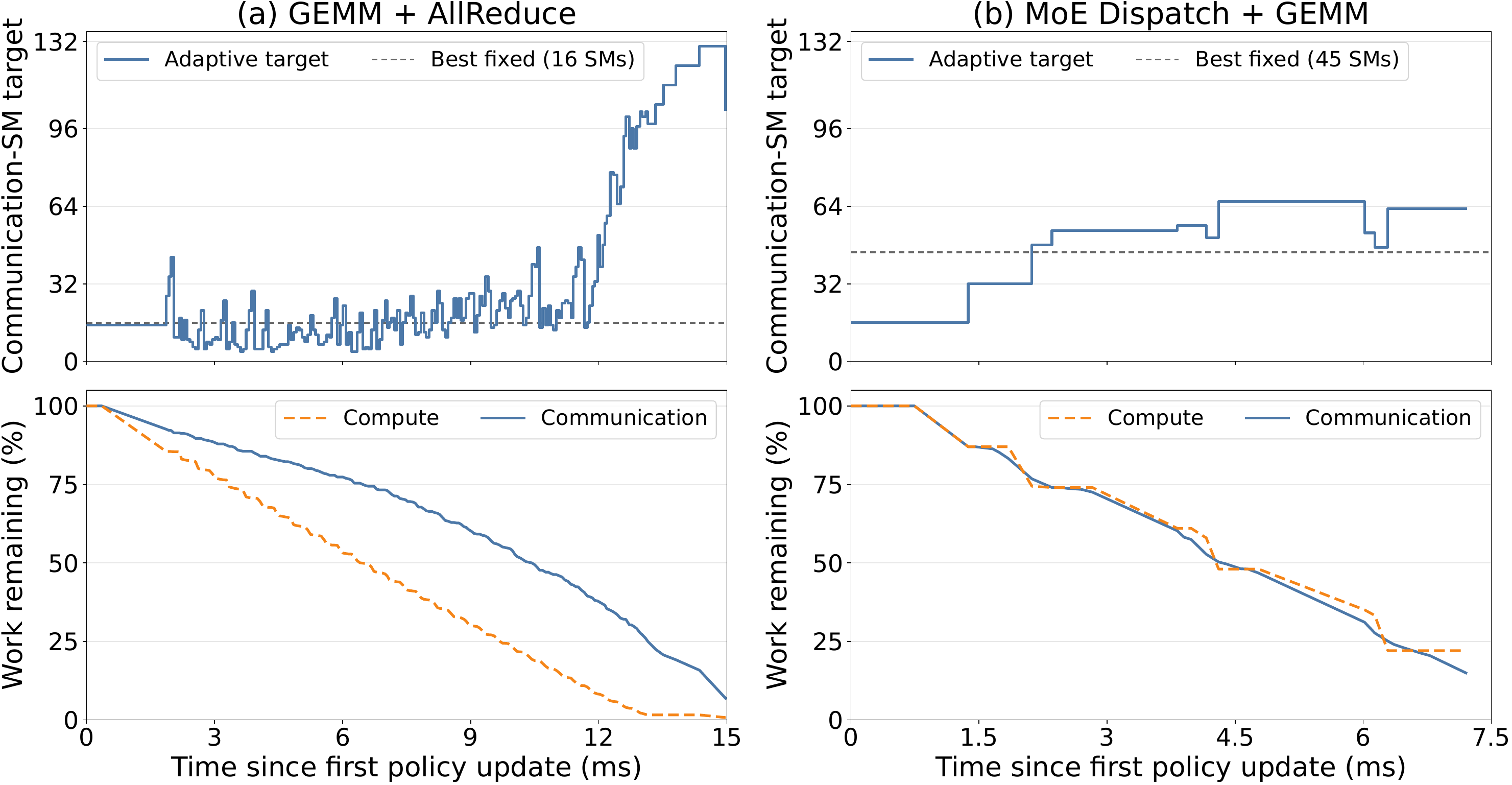}
\caption{\textbf{Adaptive SM partition tuning}. (a) GEMM+AllReduce ($M=N=32$K, $K=4$K) shifts its target toward communication as computation nears completion. (b) MoE Dispatch+GEMM (128K total tokens) increases the dispatch-SM target while computation and communication make similar relative progress. Gray dashed lines mark the best fixed partitions (16 and 45 SMs respectively). }
\label{fig:adaptive-allocation}
\end{figure}

\paragraph{Static sweep.}
We sweep communication allocations from 1 or 2 to 64 SMs in powers of two.
\Cref{fig:adaptive-sweep} normalizes each curve to its minimum measured latency. The preferred allocation ranges from 2 to 64 SMs across workloads. Too few communication SMs can stall computation or leave a large communication backlog: the worst measured AllReduce SM partition is approximately $25\times$ slower than the best. Excess communication allocation reduces compute resources. Static sweeps recover good configurations, but the changing minima show why a single choice does not transfer across kernels and sizes.

\paragraph{Adaptive controller.}
The star markers in \cref{fig:adaptive-sweep} show the controller's latency for each size. Without per-shape tuning, it achieves a geometric mean of 1.18$\times$ across 21 configurations compared to the static best fixed SM partition.

\paragraph{Intra-kernel SM partition adaptation} \Cref{fig:adaptive-allocation} illustrates how the allocation target responds to remaining work for GEMM+AllReduce and MoE Dispatch+GEMM. For GEMM+AllReduce, since compute tasks finish early, all SMs are dedicated to communication towards the end to finish the remaining work..

\section{Related Work}
\label{sec:related}

\paragraph{Kernel frameworks and distributed fusion.}
CUTLASS~\cite{cutlass}, Triton~\cite{tillet2019triton}, and ThunderKittens~\cite{spector2025thunderkittens} provide tiled GPU computation abstractions; \system uses ThunderKittens for its compute blocks. ParallelKittens~\cite{sul2025parallelkittens} extends these abstractions with intra-node communication and synchronization. FLUX~\cite{chang2024flux} fuses communication and readiness checks into GEMM, including inter-node writes through NVSHMEM. TileLink~\cite{zheng2025tilelink}, Triton-distributed~\cite{zheng2025tritondistributed}, and Mercury~\cite{guan2025mercury} expose computation and communication to compiler optimization. \system uses explicit schedules within persistent kernels to coordinate NVLink transfers, RDMA chunks, and SM roles.

\paragraph{Overlap scheduling and SM allocation.}
Megatron-LM~\cite{narayanan2021megatron} and Transformer Engine~\cite{transformerengine} overlap communication with computation across operations. Work decomposition~\cite{wang2023overlap} and Centauri~\cite{chen2024centauri} expose smaller scheduling units for dependent operations. CoCoNet~\cite{jangda2022coconet} supports compiler transformations for fusion and overlap, while T3~\cite{pati2024t3} uses hardware tracking and triggering. NanoFlow~\cite{zhu2025nanoflow} jointly selects batch decomposition and GPU resource allocation; COMET~\cite{zhang2025comet} selects profiled compute--communication configurations using workload metadata. \system specifically targets inter-node fused kernels and updates its target SM partition during kernel execution using measured progress and remaining work (\cref{sec:adaptive}).

\paragraph{Communication interfaces and topology.}
NVSHMEM~\cite{nvshmem} supports device-side one-sided operations through GPUDirect Async (IBGDA)~\cite{ibgda} or host-assisted GPU-initiated communication; its libfabric backend supports EFA~\cite{nvshmemtransports}. MSCCL++~\cite{hwang2026mscclpp} offers peer-memory access and proxy-mediated networking; NCCL's device API and GIN~\cite{nccldeviceapi,hamidouche2025ncclgin} support device communication with direct and proxy network backends. UCCL-Tran~\cite{zhou2025uccl} places transport control on CPUs, and UEP~\cite{mao2025ucclep} uses GPU-issued commands and host proxies for portable expert-parallel communication. \system integrates a compact command queue and direct verbs implementation with fused schedules on InfiniBand and EFA. TACCL~\cite{shah2023taccl} synthesizes topology-aware collectives; \system uses hierarchical transfer schedules constrained by tile readiness.

\paragraph{Persistent kernels and megakernels.}
The Llama megakernel~\cite{spector2025megakernel} and MPK~\cite{cheng2025mpk} use persistent execution to schedule model operations, with MPK supporting multi-GPU inference. \system focuses on computation and communication within each distributed kernel.

\section{Conclusion}
\label{sec:conclusion}

\system is a library of \textbf{multi-GPU, multi-node fused kernels that overlap computation and communication at tile granularity}. \system adopts SM specialization, hierarchical data movement, and portable host-assisted GPU-initiated communication, which enables the same kernels to coordinate NVLink and inter-node RDMA on InfiniBand and EFA. We implement five kernels for tensor, sequence, and expert parallelism. On two 16-GPU H200 clusters, \system achieves speedups of up to $1.72\times$ on GEMM+AllReduce and $1.88\times$ on Ring Attention over the unfused baselines. The intra-node adaptive mode complements this design by dynamically tuning compute and communication SM allocations, reducing overhead of tuning kernels for each kernel type and input shapes.

\section*{Acknowledgement}

We would also like to thank Chon Lam Lao, Xingyu Xiang
for helpful discussions. This work is in part supported by gifts from Accenture, AMD, Anyscale, Broadcom, Cisco, Google, IBM, Intel, Intesa Sanpaolo, Lambda, Lightspeed, Mibura, Microsoft, NVIDIA, Samsung SDS, and SAP. 
Costin Raiciu was partly funded by HRIA (project no. 351416). Yihan Zhang, Shuang Ma, and Yang Zhou are supported by NSF Grant 2552193.

\bibliographystyle{plainnat}
\bibliography{paper}

\end{document}